\documentclass[aps,prb,showpacs,twocolumn,groupedaddress]{revtex4}
\usepackage{amsmath}
\usepackage{amssymb}
\usepackage{graphicx,color}
\usepackage{hyperref}

\begin{document}

\preprint{1.0}

\title{Full Counting Statistics of Quantum Point Contact with Time-dependent Transparency}


\author{Jin Zhang}
\author{Y. Sherkunov}
\author{N. d'Ambrumenil}
\author{B. Muzykantskii}
\affiliation{Department of Physics, University of Warwick, Coventry, CV4 7AL, United Kingdom}


\date{\today}

\begin{abstract}
We analyse the zero temperature Full Counting Statistics (FCS) for the charge transfer across
a biased tunnel junction.
We find the FCS
from the eigenvalues of the density matrix of outgoing states of one lead.
In the general case of a general
time-dependent bias and time-dependent transparency
we solve for these eigenvalues numerically.
We report the FCS for the case of
a step pulse applied between the leads and a constant barrier transparency
(this case is equivalent to Fermi edge singularity problem). We have also
studied combinations of a time-dependent barrier transparency
and biases between the leads. In particular we look at protocols which
excite the minimal number of excitations for a given charge transfer (low noise electron source)
and protocols which maximise entanglement of charge states.
\end{abstract}


\maketitle

\section{Introduction}\label{sec:intro}
Charge fluctuations in mesoscopic devices are increasingly important
as the devices become smaller. At low temperatures, the statistics
of these fluctuations is determined by quantum effects. Attention
has focused on the full distribution of probabilities $P_n$ for the
transfer of $n$ charges from one part of a system to another---the
so-called Full Counting Statistics. \cite{LevitovJETP93} In
fermionic systems, the quantum nature of the system, and how it is
driven by external stimuli, manifests itself, even for
non-interacting fermions, in the current-current correlation
function and the higher order correlation functions which are
becoming increasingly accessible to experiment.
\cite{Gustavsson06,Toshimasa06,Gustavsson08}

Most theoretical work has concentrated on the simplest possible
device, namely a tunnel junction between two 1D leads.
\cite{LesovikLeeLevitov96,IvaLL97,Nazarov01,MA03,dAMuz05,KeelingKlichLev06,VanevicNazBelz07,Abanov08}
It has been shown that, if a bias pulse, $V(t)$, is applied across a
junction with fixed transparency at low temperature, the tunneling
processes induced by the pulse are combinations of two elementary
types of event called uni-directional and
bi-directional\cite{VanevicNazBelz07}. It is also known that the
statistics of the transfer of charge induced by such pulses depend
strongly on the driving protocol. In contrast to a general shape of
ac pulse which leads to an indefinite number of electronic
excitations in the leads, an optimal ac signal, which is composed of
overlapping Lorentzian pulses, has been found to excite a strictly
finite number of excitations per cycle and to bring the noise down
to dc levels. \cite{IvaLL97} The minimal excitation states (MES),
\cite{KeelingKlichLev06} created by such optimal pulses, offer the
prospect of being able to generate signals with well-defined charge
transfer down to the level of single electron emission. Coupled with
the high Fermi velocity in electronic systems there is the prospect
of rapid solid-state information transfer at a level useful for
quantum information processing.
\cite{NederNature,Beenakker04,BeenakkerTitov05,Trauzettel06,Samuelsson09}

Another area where the FCS have been studied is that of quantum
pumps. These can lead to the pumping of electrons from one side of a
tunneling barrier to the other. Several schemes for operating a
tunnel junction, which can lead to the transfer of charges
\cite{AndreevKamenev2000,BeenakkerTitov05} and produce entangled
electron-hole pairs in separate leads, have been proposed.
Samuelsson and B\"uttiker have proposed an orbital-entangler, which
works with quantum Hall edge states. \cite{Samuelsson04} Accurate
control of the transparency may allow the generation and control of
flying qubits, while Beenakker \textit{et al.} have shown that such
an electronic entangler based on a biased point contact could reach
the theoretically maximum efficiency of $50$\%. \cite{Beenakker05}

If the quantum effects are not to be obscured by thermal noise, the
temperature $1/\beta$ must be low enough that $t_f < \beta$, where
$t_f$ is the measurement time (or inverse repeat frequency for an ac
measurement). Working at temperatures around $10$mK would require
operating at frequencies around 200MHz, and this is the temperature
and frequency regime used in some experiments. \cite{Feve07}
However, even at zero temperature the so-called equilibrium noise is
present and diverges logarithmically with the inverse repeat
frequency or measurement time $t_f$. This equilibrium noise is
present in both the proposed MES protocol for generating charge
transfer and the protocol for the optimal electronic entangler.

Here we develop our approach to calculating the FCS for a tunnel
junction\cite{Sherkunov08} and examine protocols which we proposed
for suppressing the equilibrium noise both in the case of the charge source and of the entangler.
\cite{Sherkunov09} We show how to solve for the FCS in the general
case of fully time-dependent barrier profile with dynamic bias
pulses applied between the leads. We compute the resulting FCS and
induced entanglement entropy for a number of profiles and, in
particular, those close to optimal (in the sense that they have low
noise in the case of electron sources or maximum entanglement). Our
approach is motivated partly by the result of Abanov and Ivanov
(AI), \cite{Abanov08} who on quite general grounds deduced
constraints on the analytic properties of the characteristic
function (generating function for the probability distribution
$P_n$). Although they have recently argued that at any temperature,
the counting statistics can be regarded as generalized binomial
statistics in which electrons scatter off the barrier with some
effective transparency independently, \cite{Abanov09} our results
are all for zero temperature.

The paper is organized as follows. In Sec. \ref{sec:method} we
review FCS and we give a short alternative derivation of the AI
formula. We then discuss the two `standard' special cases: the
biased contact with fixed transparency, and a contact with modulated
barrier transparency. We establish a mapping between these two cases
and use this to simplify the derivation of the FCS for these two
cases and to explore the relation to the Fermi Edge Singularity (FES)
problem. In Sec. \ref{sec:general} we describe a general purpose
numerical procedure to solve for the general case inaccessible to
analytical techniques. We use this to compute the effects of
deviations from the ideal voltage pulses, which lead to minimal
noise in the charge transferred across a tunnel barrier when
operated as an electron source, and to study protocols close to
optimal for the generation of electron entanglement. Concluding
remarks can be found in Sec. \ref{sec:conclusion}.

\section{Full Counting Statistics}\label{sec:method}

\begin{figure}[!ht]
    \begin{center}
        \includegraphics[width=0.25\textwidth]{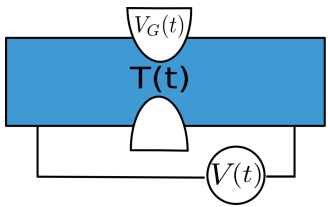}
    \end{center}
    \caption{
    \label{fig:qpc}
    (Color online) A quantum point contact
     with time-dependent bias voltage $V(t)$ applied on the right lead.
    The transparency of the QPC, $T(t)$, is controlled by the gate voltage $V_G(t)$.\cite{Gershon08}}
\end{figure}

We consider a quantum point contact (QPC) at zero temperature with
time-dependent transparency, $T(t)$, connecting two single channel
ballistic conductors, as illustrated in Fig. \ref{fig:qpc}. We
assume there is no inelastic scattering inside the QPC or the leads.
The two leads (assumed identical) are disconnected initially and
contain non-interacting electrons in their respective groundstates
$|0\rangle$. Electrons in the disconnected leads are described by
the Hamiltonian
$H_0=\sum_j\varepsilon_j\mathbf{c}_j^\dagger\mathbf{c}_j$, where we
assume the system has a discrete energy spectrum. The electron
creation operators $\mathbf{c}^\dagger$ have been written as a
vector $\mathbf{c}^\dagger=(c_L^\dagger, c_R^\dagger)$, where
$c_{L,R}^\dagger$ is the creation operator for states in the left
and right lead respectively. The groundstate of a single lead
$|0\rangle$ is the Fermi sea,
$|0\rangle=\prod_{\varepsilon_j<\mu}c^\dagger_j|\rangle$, where
$|\rangle$ is the ``true'' vacuum and $\mu$ is the Fermi energy.
Where necessary, we will assume a cut-off of order the Fermi energy.

The two Fermi seas are initially uncoupled. Usually it is assumed
\cite{MA03} that the tunneling barrier is lowered at time
$t=0$, allowing electrons to tunnel between the two leads, and is
restored to fully reflective after a measurement time, $t_f$. In
general, the evolution of outgoing states should be described by
solving the fully time-dependent Hamiltonian $H(t)=H_0+H'(t)$, with
\begin{equation}
H'(t)=\sum_{j,j'}\mathbf{c}_j^\dagger
\mathbf{M}(t,\varepsilon_j,\varepsilon_{j'})\mathbf{c}_{j'}.
    \label{eqn:ham}
\end{equation}
The matrix $\mathbf{M}(t)=0$ for $t<0$ and $t>t_f$. However, if the
scattering potential varies slowly on the scale of the Wigner delay
time $\tau_d \ll \tau_W \sim S^{-1}\frac{\partial S}{\partial E}$,
with $\hbar= e= 1$,
the properties of the system can be
determined from the instantaneous value of scattering matrix $S(t)$
evaluated on states at the Fermi energy
\begin{equation}
    S(t,E=\mu) = \left(\begin{array}{cc}
           B(t)     &    A(t)\\
        -A^\ast(t)  &  B^\ast(t)
    \end{array}\right).
    \label{eqn:smat_general}
\end{equation}
Here $A(t)$ and $B(t)$ are time-dependent transmission and
reflection amplitudes and are determined by both the QPC gate
voltage and the bias voltage applied between the leads. This relates
eigenstates $c_{j}$ of $H_0$, which we separate into incoming
$a_{j}$ and outgoing states $b_j$, via
\begin{equation}
    \begin{pmatrix}b_{L}(t)\\b_{R}(t)\end{pmatrix} = S(t)\begin{pmatrix}a_{L} \\a_{R}\end{pmatrix}.
    \label{eqn:atob}
\end{equation}

We are interested in the distribution $P_n$, which is the
probability that there is a net transfer of $n$ charges from the
left to the right lead during the measurement period $0<t<t_f$. A
convenient way of characterizing the Full Counting Statistics (FCS),
$P_n$, is via the function $\chi(\lambda)$:
\begin{equation}
    \chi(\lambda) = \sum_{n=-\infty}^\infty P_n e^{i\lambda n}.
    \label{eqn:cf}
\end{equation}
The current, noise and higher order cumulants, $\langle\langle
Q^m\rangle\rangle$, can be computed from $\chi(\lambda)$:
$\langle\langle Q^m\rangle\rangle = \frac{\partial^m
\ln(\chi)}{\partial (i\lambda)^m}|_{\lambda=0}$, where $m$ is the
order of the cumulant. The formula for the FCS
is:\cite{LevitovJETP93,Klich02Book,MA03}
\begin{equation}
    \chi(\lambda)=\det(1+n(S^\dagger e^{i\lambda L}S e^{-i\lambda L}-1)),
    \label{eqn:levitov}
\end{equation}
where $n$ is the number operator for the fermions. The matrix $L$
projects onto states in the left lead:
$\begin{pmatrix}1&0\\0&0\end{pmatrix}$ in lead space. The matrix
inside the determinant is infinite dimensional in the energy or time
domain and $2\times 2$ in lead space. Because of the infinite
dimensionality of the space of states in (\ref{eqn:levitov}),
careful regularization of the formula is required.
\cite{Klich02Book,MA03,Avron2007} For example, at very high energies
$\varepsilon\rightarrow+\infty$, $n=0$, the argument of the
determinant approaches the identity and the contributions remain
finite and computable. However, when
$\varepsilon\rightarrow-\infty$, $n=1$, and the matrix has the
asymptotic form $S^\dagger e^{i\lambda L}S e^{-i\lambda L}$, which
makes the determinant ill-defined for an infinitely deep Fermi sea.

A simple and correctly regularized approach to the computation of
$\chi(\lambda)$ works with the density matrix for the
\textit{outgoing} states in one of the leads. The density matrices
for incoming states in both leads can be written
$n^{in}=\langle0|a^\dagger_j
a_j|0\rangle=n_j=\theta(\mu-\varepsilon_j)$, which is the Fermi
distribution function at zero temperature. Fourier transformed to
the time domain, the density matrix has the form $n^{in}(t,t') = \int
d\varepsilon \theta(\mu-\varepsilon)
e^{i(t-t')(\varepsilon-\mu)}=\frac{i}{2\pi}\frac{1}{t-t'+i0}$. The
density matrix of the outgoing states in, say, the left lead
$n^{out}(t,t')$ can be obtained from (\ref{eqn:atob}):
\begin{eqnarray}
    \nonumber
    n^{out}(t,t') &=& \langle b_L^\dagger(t)b_L(t')\rangle\\
    &=& B^\ast(t) n^{in}_L(t,t')B(t')+A^\ast(t) n^{in}_R(t,t')A(t').
    \label{eqn:nout}
\end{eqnarray}
In the second equation, we have used the fact that terms like
$\langle a^\dagger_La_R\rangle$ are zero, because the incoming
states between different leads are uncorrelated.

From $n^{out}$, it is possible to compute the cumulants directly.
For example, the second cumulant (noise) is
\cite{LeeLevitov93}
\begin{eqnarray}
    \nonumber
    \langle\langle Q^2\rangle\rangle &=&
    2\iint dtdt'n^{in}(t,t')\left[1-n^{in}(t',t)\right]\\
    & &\left[|A(t)|^2|A(t')|^2+A^\ast(t)B(t)A(t')B^\ast(t')\right].
    \label{eqn:noise}
\end{eqnarray}
In the case, when the barrier transparency is switched on and off
with $A(t)=A_0=const$ for $0<t<t_f$ and $A(t)=0$ otherwise, and with
no bias voltage applied between the leads, the so-called equilibrium
noise is obtained from the integral in (\ref{eqn:noise}):
$\langle\langle Q^2\rangle\rangle=\frac{A_0^2}{\pi^2}\log t_f\xi$.
Here $\xi$ is the commonly used ultraviolet energy cut-off of the
order of Fermi energy. The logarithmic term is present for almost
all profiles and not just abrupt switching. It was found, for
example, for the case of a Gaussian switching
profile.\cite{Moskalets07}

The eigenvalues, $n_j$, of $n^{out}$ allow for the direct
computation of the FCS. We consider first a simple case, where only
one eigenvalue changes to $n_j$ and all other eigenvalues are
unchanged. Since all eigenvalues initially are either 0 or 1, we
need only to consider a change $0\rightarrow n_j$ or $1\rightarrow
n_j$. In the first case, the probability of one additional
particle being transferred into state $j$ from the right lead is $n_j$, while
the probability, that
no particle is added, $(1-n_j)$. The counting
statistics follow from (\ref{eqn:cf}) and are given by
$\chi_j(\lambda)=1-n_j+e^{i\lambda}n_j$, with the average charge
transfer given by $\langle Q\rangle=\frac{\partial \ln\chi}{\partial
(i\lambda)}|_{\lambda=0}=n_j$. If the occupation changes from
$1\rightarrow n_j$, a single hole is transferred with probability
$1-n_j$ while no charge transfer takes place with probability $n_j$.
For this case, $\chi_j(\lambda)=n_j+e^{-i\lambda}(1-n_j)$, with the
average charge transfer $\langle Q\rangle=n_j-1$. The results for
the two cases can both be written
\begin{equation}
\chi_j(\lambda)=e^{i\lambda (\langle
Q\rangle-n_j)}[1+(e^{i\lambda}-1)n_j]. \label{eqn:chi_j}
\end{equation}

Since we work in the basis where $n^{out}$ is diagonal, the result
for $\chi(\lambda)$ for the general case is simply a product over
the factors, $\chi_j(\lambda)$. Taking account of all possible
processes, we arrive at the formula
\cite{Sherkunov09,Abanov08,Klich02Book,Avron2007},
\begin{equation}
\chi(\lambda) = e^{i\lambda \langle Q\rangle}\prod_{j=1}^N
e^{-i\lambda n_j}[1+(e^{i\lambda}-1)n_j].
    \label{eqn:ivanov}
\end{equation}
It is correctly regularized as states unaffected by the perturbation
contribute a factor $1$ to $\chi(\lambda)$. It is also well suited
to direct numerical calculation.

In the following sections, we will discuss two special cases where
the FCS can be obtained analytically. We rederive the known
results for these cases by working directly with the density matrix,
$n^{out}$. Then we show how the FCS for the general case are
easy to obtain by diagonalizing $n^{out}$ numerically. To facilitate
the interpretation of (\ref{eqn:ivanov}), we choose a specific form
for the scattering matrix $S(t)$, and assume that the transmission
and reflection amplitudes of the barrier $A(t)$ and $B(t)$
controlled by the QPC are \textit{real}. If a bias voltage $V(t)$ is
applied between the leads, its effect is to introduce an additional
phase difference between the states in the two leads given by the
Faraday flux $\psi(t)=\frac{e}{\hbar}\int_0^t V(t')dt'$. We
incorporate this effect via a gauge transformation applied to states
in the right lead: $a_R\rightarrow a_Re^{i\psi(t)}$. The resulting
scattering matrix $S(t)$ is:
\begin{equation}
    S(t) = \left(\begin{array}{cc}
        B(t)  &  A(t)e^{i\psi(t)}\\
        -e^{-i\psi(t)}A(t) &  B(t)
    \end{array}\right).
    \label{eqn:smat}
\end{equation}

\subsection{Bias-voltage applied between the leads}\label{ssec:voltage}
The case of a bias voltage pulse, $V(t)$, applied across a barrier
with fixed transmission amplitude, $A$, for $0<t<t_f$ has been well
studied. \cite{LevitovJETP93,LesovikLeeLevitov96,IvaLL97} It has
been shown that charge transfer processes at zero temperature are
made up of combinations of two elementary events called
uni-directional and bi-directional. \cite{IvaLL97,VanevicNazBelz07}
The uni-directional event relates to a single charge transfer
process associated with the dc component of the bias voltage $V(t)$.
A single state is occupied on one side of the barrier and not on the
other. This leads to the possibility of the transfer of charge
across the barrier but only in one direction. Unidirectional events
contribute to the average current as well as higher order cumulants.
Bi-directional events are the consequence of the ac component of
$V(t)$, and relate to the excitation of equal numbers of particles
and holes. These can both be transferred or reflected at the barrier
so that charge can be transferred in either direction. No average
current is generated in this case and only even cumulants are
non-zero. The generic formula for the FCS for charge transfer is
\cite{VanevicNazBelz07}
\begin{equation}
    \chi(\lambda) = \prod_{i=1}^{N_u}(R+Te^{i\kappa\lambda})
    \prod_{j=1}^{N_b}[1+RT\sin^2\frac{\alpha_j}{2}(e^{i\lambda}+e^{-i\lambda}-2)].
    \label{eqn:vnb}
\end{equation}
$T=|A|^2$ is the barrier transparency and $R=1-T$. $N_u$ and $N_b$
correspond to the total number of uni- and bi-directional events.
$\kappa=\pm 1$ depending on the polarity of the voltage pulse. The
angles $\alpha_j/2$ determine the probability of exciting a single
particle-hole pair in a bi-directional event. $N_u$, $N_b$ and
$\alpha_j/2$ can be computed\cite{VanevicNazBelz07} by diagonalizing matrix $h\tilde{h}$,
where $h$ and $\tilde{h}$ are defined as $h=2n-1$ and
$\tilde{h}=UhU^\dagger$, with $U(t)=e^{i\psi(t)}$.

We can understand the form of (\ref{eqn:vnb}) by considering the
density matrix of outgoing states, which in the
case of constant transparency has the form
\begin{equation}
    n^{out}(t,t') = Rn^{in}_L(t,t')+Te^{i\psi(t)}n^{in}_R(t,t')e^{-i\psi(t')}.
    \label{eqn:nout_voltage}
\end{equation}
We assume that the measurement time is short enough that we can
ignore the equilibrium noise contribution, which is a
logarithmically divergent function of the measurement time $t_f$.
\cite{Sherkunov09} The equilibrium noise is associated with
fluctuations in the number of particles in the left or right lead
and occurs even in the absence of an applied voltage.

The dc component of the voltage pulse, associated with non-zero
Faraday flux $\psi$, generates additional occupied particle (or
hole) states in the right lead when compared with the incoming
states in the left lead. The corresponding particle (or hole), after
impinging on the barrier, will tunnel across with probability
$T=A^2$ or be reflected with probability $R=1-T$. This gives rise to
so-called uni-directional events. The eigenvalue of the density
matrix for outgoing states in the left lead is then
$n_j=T$, with an average charge transfer from right to left of
$\langle Q \rangle = T$ if $j$ relates to a state above the Fermi
energy, or $n_j=1-T$ and $\langle Q \rangle=-T$ if $j$ relates to a
state below the Fermi energy. Inserting this in (\ref{eqn:ivanov}),
gives $\chi_u(\lambda)=R+Te^{i\kappa\lambda}$, where $\kappa=\pm1$
is determined by the type of transferred charge (particle or hole).

An example of unidirectional events is provided by the so-called
minimal excitation states (MES). These excite a number of particles
(or holes) with minimum noise. \cite{IvaLL97,KeelingKlichLev06} The
corresponding voltage pulse $V(t)$ is a sum of Lorentzian
pulses: $V(t)=\sum_j^N\frac{2\tau_j}{(t-t_j)^2+\tau_j^2}$, where
$t_j$ and $|\tau_j|$ determine the center and width of the $j$th pulse
respectively. The unitary
transformation in  (\ref{eqn:smat}) is \begin{equation}
    e^{i\psi(t)} = \prod_j^N\frac{t-t_j-i\tau_j}{t-t_j+i\tau_j}.
    \label{eqn:lorentzian}
\end{equation}
Choosing the signs of the $\tau_j$ to be the same leads to $N$
unidirectional events. For $N=1$, the pulse in
(\ref{eqn:lorentzian}) generates a single uni-directional event with
one particle (electron or hole, depending on the polarity of the
pulse) passing through the barrier with probability $T$ giving
$\chi(\lambda) = R+Te^{ik\lambda}$. For $N=2$, with the polarities
of the two pulses (set by the signs of the $\tau_j$) the same, the
two uni-directional events are combined and
$\chi(\lambda)=(R+Te^{i\kappa\lambda})^2$ irrespective of the
relative widths ($|\tau_j|$) or positions ($t_j$).

The ac component of the voltage pulse, associated with zero total
Faraday flux, gives rise to the so-called bi-directional
events---the simplest example of which is given by a pulse of the
type (\ref{eqn:lorentzian}) with $N=2$ and $\tau_1 \tau_2 < 0$. The
operator $h\tilde{h}$ characterizes the differences between the
states in the two leads following the application of the pulse. All
states, which are either both occupied or both empty in the two
leads and which therefore contribute a factor 1 to $\chi(\lambda)$,
are eigen states of $h\tilde{h}$ with eigenvalue 1 ($\alpha_j=0$).
All other eigenvalues occur in pairs and are equal to $e^{\pm
i\alpha_j}$. \cite{VanevicNazBelz07} In the basis of the unperturbed
states of either of the leads, the state described by $\tilde{h}$
contains an admixture of the unperturbed state and independent
particle and hole excitations in each of the $2$ dimensional
subspaces of the basis (labeled by $j$), in which $\tilde{h}$ is
block diagonal. Its eigenvalues are $e^{\pm i\alpha_j}$. Each of
these will lead to eigenvalues in $n^{out}$ of $n_j$ and $(1-n_j)$, one
associated with the particle excitation and one with the hole.

The contribution to
$\chi(\lambda)$ from each of these bi-directional events
(corresponding to the different $j$) is the product over the two
factors of the type in (\ref{eqn:chi_j}), one with eigenvalue $n_j$
($\langle Q\rangle=n_j$) and one with eigenvalue $1-n_j$ ($\langle
Q\rangle=n_j-1$):
$\chi_b(\lambda)=\left(1+(e^{i\lambda}-1)n_j\right)\times\left(1+(e^{-i\lambda}-1)(1-n_j)\right)
=1+n_j(1-n_j)(e^{i\lambda}+e^{-i\lambda}-2)$. To make a connection
between $n_j$ and rotation angle $\alpha_j/2$, we use the known
result\cite{VanevicNazBelz07}
$h=\begin{pmatrix}0&1\\1&0\end{pmatrix}$ and
$\tilde{h}=\begin{pmatrix}0&e^{-i\alpha_j}\\e^{i\alpha_j}&0\end{pmatrix}$
in the eigenbasis of $h\tilde{h}$. Substituting $h$
and $\tilde{h}$ into (\ref{eqn:nout_voltage}) and diagonalizing
$n^{out}$ explicitly, we obtain
$n_j=\frac{1}{2}\pm\frac{1}{2}\sqrt{1-4RT\sin^2\frac{\alpha_j}{2}}$
and
\begin{equation}
    n_j(1-n_j)=RT\sin^2\frac{\alpha_j}{2}.
    \label{eqn:nj_alf}
\end{equation}

After taking the product over all events labeled by $j$, and adding
in the contribution of the uni-directional events, we recover
(\ref{eqn:vnb}).
For a given voltage pulse between the leads and corresponding
unitary transformation $U(t)$, the $\alpha_j/2$ can be thought of as
the rotation angles of the ground state associated with $U(t)$ and
are found by diagonalizing the $h\tilde{h}$.
\cite{VanevicNazBelz07,Sherkunov08} The rotated state is an
admixture of the original state, with probability $\cos^2
\frac{\alpha_j}{2}$, and the state with an added particle and hole,
with probability $\sin^2 \frac{\alpha_j}{2}$. The factor
$1+RT\sin^2\frac{\alpha_j}{2}(e^{i\lambda}+e^{-i\lambda}-2)$ is the
weighted average of the result for the unperturbed state
(contribution 1 with weight $\cos^2\frac{\alpha_j}{2}$) and for the
state with an added particle and hole (contribution
$(R+Te^{i\lambda})(R+Te^{-i\lambda})$ with weight
$\sin^2\frac{\alpha_j}{2}$).\cite{Sherkunov08}

\subsection{Barrier with modulated Transparency}\label{ssec:barrier}
Another case for which results in closed form have been reported is
that of a time-dependent barrier between two leads at the same
chemical potential. \cite{AndreevKamenev2000, Sherkunov09} The
problem can be mapped onto a special case of a voltage biased
time-independent barrier with constant transmission and reflection
amplitudes. \cite{Sherkunov09} This mapping becomes explicit once
the problem is approached via the density matrix of the outgoing
states. In the absence of a bias the scattering matrix in
(\ref{eqn:smat}) simplifies to $S(t) = \left(\begin{array}{cc} B(t)
& A(t)\\-A(t) & B(t)\end{array}\right)$ and the density matrix of
outgoing states becomes
$n^{out}(t,t')=B(t)n(t,t')B(t')+A(t)n(t,t')A(t')$. Introducing
$e^{i\phi(t)}=B(t)+iA(t)$ (we are still assuming that $A$ and $B$
are real), we insert $e^{i\phi(t)}$ into $n^{out}$ and eliminate $A$
and $B$. We obtain
$n^{out}=\frac{1}{2}(e^{i\phi}n^{in}e^{-i\phi}+e^{-i\phi}n^{in}e^{i\phi})$.
Here we have used the fact that $n^{in}_L=n^{in}_R=n^{in}$.

Since a unitary transformation on $n^{out}$ does not affect its
eigenvalues, $n_j$, we can study
\begin{equation}
e^{-i\phi}n^{out}(t,t')e^{i\phi} =\frac{1}{2}n^{in}(t,t')
+\frac{1}{2}e^{2i\phi(t)}n^{in}(t,t')e^{-2i\phi(t')}.
\label{eqn:nout_barrier}
\end{equation}
The relations (\ref{eqn:nout_barrier}) and (\ref{eqn:nout_voltage})
have the same structure. The FCS of a system with modulated barrier
transparency without a bias between the leads are therefore
equivalent to those for a system with bias voltage applied across a
barrier with constant transmission and reflection amplitudes
$A=B=\frac{1}{\sqrt{2}}$, and Faraday flux $\psi=2\phi$. The FCS for
a modulated barrier transparency can therefore be obtained from
(\ref{eqn:vnb}). In addition, other concepts developed to understand
the bias voltage case carry over to the modulated barrier case.
These include the geometrical interpretation of the FCS,
\cite{Sherkunov08} as well as the MES, \cite{KeelingKlichLev06}
which, when implemented as a modulation profile of the barrier leads
to the optimal entangler of electron hole pairs.\cite{Sherkunov09}

To calculate $\chi(\lambda)$ for the case of the modulated barrier
from (\ref{eqn:vnb}), we need, as before, to diagonalize the matrix
$h\tilde{h}=he^{2i\phi}he^{-2i\phi}$ and compute the angles
$\alpha_j/2$ from its eigenvalues. Since no bias voltage is applied
between the leads, the system is completely symmetric in lead space.
As a result only bi-directional events (leading to no net average
charge transfer) can occur and all the eigenvalues of
$he^{2i\phi}he^{-2i\phi}$ come in pairs. The characteristic function
$\chi(\lambda)$ is then given by
\begin{eqnarray}
    \nonumber
    \chi(\lambda) &=& \prod_{j=1}^{N}\left(1+\frac{1}{4}\sin^2\frac{\alpha_j}{2}
                    (e^{i\lambda}+e^{-i\lambda}-2)\right)\\
    &=& \prod_{j=1}^{N}\left(1-\sin^2\frac{\alpha_j}{2}\sin^2\frac{\lambda}{2}\right),
    \label{eqn:cf_barrier}
\end{eqnarray}
where $N$ is the number of paired eigenvalues determined from
$he^{2i\phi}he^{-2i\phi}$.

To illustrate the value of this mapping, consider the case in which
the transparency of a barrier is subjected to a sinusoidal
modulation: $A(t)=\sin\omega t$ and $B(t)=\cos\omega t$. If the
total number of cycles is large enough, the contribution from the
equilibrium noise (proportional to $\log t_f\xi$) can be neglected.
This problem can be mapped to a case of a barrier with constant bias
voltage with transparency $T=1/2$ and $\psi = \int^t V(t') dt' =
2\phi = 2\omega t$ corresponding to a constant dc bias with $V=
2\omega$.

\begin{figure}[t]
    \begin{center}
        \includegraphics[width=0.4\textwidth]{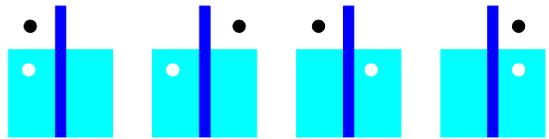}
    \end{center}
    \caption{
    \label{fig:sinu}
    (Color online) Four possible outcomes per period from a sinusoidally modulated
    barrier transparency. All four outcomes are equally likely.
    The particle-hole pair can be on either side
    of the barrier or as an entangled pair of particle on one side and hole on the other.}
\end{figure}

To obtain the FCS, we need to diagonalize matrix $he^{2i\omega
t}he^{-2i\omega t'}$. As this corresponds to a constant dc bias
problem involving two uni-directional events per period,
(the phase changes by $4\pi$ per period), there are
two eigenvalues different from 1 and both are equal to -1. The
polarity (particle or hole) of transferred charge can be inferred
from the requirement that there is no net average charge transfer.
Hence we conclude that the sinusoidally modulated transparency case
is equivalent to one electron and one hole impinging in a single
period on a barrier with transmission $T=1/2$. The two processes are
independent since both correspond to uni-directional events in the
equivalent bias voltage problem. The FCS for constant bias voltage
case are known to be given by
$\chi_0(V,\lambda)=(R+Te^{i\kappa\lambda})^{\frac{t_fV}{4\pi}}$,
with $\kappa=\pm1$ depending on the polarity of the applied voltage.
\cite{LeeLevitov93,MA03} The corresponding FCS for the sinusoidally
modulated transparency case is a combination of two factors $\chi_0$
one with $\kappa = 1$ and one with $\kappa=-1$. This gives
\begin{eqnarray}
    \nonumber
    \chi(\lambda) &=& \chi_0(\omega,\lambda)\chi_0(\omega,-\lambda)
\\ &=&
\left(\frac{1+\cos\lambda}{2}\right)^{\frac{\omega t_f}{2\pi}},
    \label{eqn:kamenev}
\end{eqnarray}
which is a result previously obtained by Andreev and Kamenev using
the Keldysh formalism. \cite{AndreevKamenev2000} A barrier operated in
this way acts as a quantum pump which excites exactly one
electron-hole pair per period, provided the logarithmically
divergent equilibrium noise is neglected. The four possible outcomes
per cycle are shown in Fig. \ref{fig:sinu}.

The result (\ref{eqn:kamenev}) is the ac version of the barrier
profile required to generate optimal electron entanglement.
\cite{Sherkunov09} The profile
\begin{equation}
e^{i\phi(t)}=B(t)+iA(t)=\frac{t-i\tau}{t+i\tau}
    \label{eqn:entangler}
\end{equation}
generates the FCS $\chi(\lambda) = (1+\cos\lambda)/2$. The use of a
quantized Lorentzian pulse of the type (\ref{eqn:entangler}) has the
advantage that the equilibrium noise is strictly absent in
(\ref{eqn:entangler}). Although the entanglement generated by
(\ref{eqn:entangler}) is between electrons and holes and therefore
not useful in quantum computation due to the charge conservation
laws,\cite{KlichLevitov0812} the protocol would be useful if
combined with another degree of freedom (either spin or orbital)
because it operates in a one-shot mode. The difficulty with its
operation is associated with the precise generation of the actual
profile. We explore the effect of possible errors in its
experimental implementation in Sec.\ref{ssec:nonop}.

\subsection{Fermi Edge Singularity}\label{ssec:fes}
We discuss here the relation between the FCS of a contact subjected
to sharp bias voltage pulses and the Fermi Edge Singularity (FES). We map
to the problem of an unbiased contact for which the result for the
FCS is known.

We consider two delta pulses with opposite signs separated by $t_f$
applied to one lead. The contact has fixed transparency $T$ and
reflectance $R$. The corresponding voltage profile is
\begin{equation}
V(t)=\frac{\psi}{2\pi}[\delta(t)-\delta(t-t_f)],
\label{eqn:delta_functions}
\end{equation}
with $\psi=const$. This pulse induces a phase shift $\psi$ on the
incoming states of, say, the left lead (measured with respect to the
right one) within time window $0<t<t_f$ and zero phase shift
otherwise. For $\psi \gg 2\pi$ and large $t_f \xi$, the noise can be
written as an expansion in $(\xi t_f)^{-1}$
\begin{equation}
    \langle\langle Q^2\rangle\rangle =
    2RT[\frac{2}{\pi^2}\sin^2\frac{\psi}{2}\ln t_f\xi+\frac{\psi}{2\pi}]+\cdots .
    \label{eqn:fes_noise}
\end{equation}
The derivation of (\ref{eqn:fes_noise}) is essentially the same as
that given by Lee and Levitov (LL) for the quantum current
fluctuations induced by a magnetic field in a metallic loop
containing a QPC.\cite{LeeLevitov93} The FCS for the problem we are
considering have been computed\cite{Sherkunov08} but only for the
case that the transparency of the barrier is low.  In this section
we address the general problem with arbitrary contact transparency.

\begin{figure}[t]
    \begin{center}
        \includegraphics[width=0.4\textwidth]{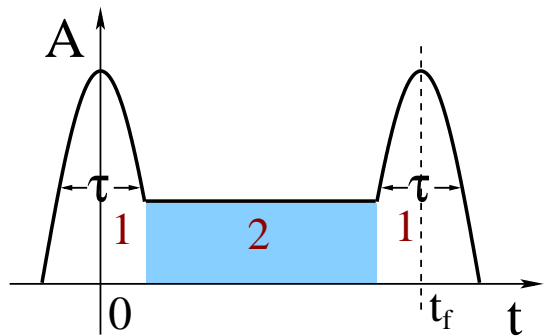}
    \end{center}
    \caption{\label{fig:connect}
    (Color online) Illustrative transmission amplitude $A(t)$ as a function of time $t$ for the
    the pulsed lead case after mapping to the equivalent barrier opening problem. The FCS are
    only well-defined if the exact time-dependence, whose limit is the delta function
    in (\ref{eqn:delta_functions}), is specified. Here we show a case where the amplitude,
    $A(t)$, actually overshoots the value $\sin (\psi/2)$. We argue that the FCS should be
    well approximated by a product over two independent contributions from the two regions
    denoted by $1$ and $2$ (shaded region).}
\end{figure}

As the two electrodes are equivalent, only bi-directional events occur in this
problem. By (\ref{eqn:vnb}), the counting statistics are
\begin{equation}
    \chi(\lambda) 
    = \prod_j[1-\sin^2\frac{\alpha_j}{2}\sin^2\frac{\tilde{\lambda}}{2}],
    \label{eqn:app_sin}
\end{equation}
where we have defined
$\sin\frac{\tilde{\lambda}}{2}=2\sqrt{RT}\sin\frac{\lambda}{2}$. The
central problem is how to compute angles $\frac{\alpha_j}{2}$. Eq.
(\ref{eqn:app_sin}) has the same form as (\ref{eqn:cf_barrier}),
which is that for a barrier with time-dependent reflection and
transmission amplitudes $B(t)+iA(t) = e^{i\psi(t)/2}$. For
$0<t<t_f$, $A(t)=\sin\frac{\psi}{2}=const$ and $A(t)=0$ if $t<0$ or
$t>t_f$.  One interesting complication in this problem is that the
FCS are only well-defined if the limit of increasingly sharp voltage
profiles, $V(t)$, leading to the delta functions in
(\ref{eqn:delta_functions}), is specified. This is because the shape
of the voltage pulse, leading to the Faraday flux change from zero
to $\psi$, affects the number of excitations introduced by the
switching process. In the mapped barrier opening problem, the
precise time-dependence of the (rapid) opening and closing of the
barrier matters. An example of a possible profile for this opening
of the barrier is shown Fig. \ref{fig:connect}.

For the case that the barrier
switching time $\tau$ is very short compared to the measurement time
$t_f$, we argue that the switching processes, which
excites high energy ($\sim 1/\tau$) particle-hole pairs, should not
interfere with the long time measurement process, which leads to the low
energy excitations, predominantly on the energy scale $\sim 1/t_f$,
expected for a Fermi Edge Singularity problem. We make the
Ansatz that the FCS can be written as a product of the types of
processes as
\begin{equation}
\chi(\lambda) = \chi_{1}(\lambda)\chi_{2}(\lambda),
\label{eqn:chi1chi2}
\end{equation}
with $\chi_2$ giving the FES contributions (which are associated
with the shaded region, shown schematically as the shaded area in
Fig. \ref{fig:connect}). $\chi_{1}$ gives the contribution
associated with the opening profile at $t=0$ and $t_f$. We focus
here on the calculation of $\chi_2$ and postpone the computation of
$\chi_{1}$ to Sec. \ref{ssec:nonop} where numerical techniques are
adopted.

The FCS of an unbiased barrier, for which the reflection amplitude is
abruptly changed from zero to the constant value
$\sin\frac{\psi}{2}$ for a duration $t_f$, is known
and was computed using the bosonization technique.
\cite{LevitovLeeLesovik96} The result can also be found by solving a
Riemann-Hilbert problem \cite{KlichLevitov09} valid even at non-zero
temperature in Appendix \ref{sec:app1}. The FCS is
\begin{equation}
    \chi_{2}(\lambda)=\exp(-\lambda^2_\ast G),
    \label{eqn:chi2_fes}
\end{equation}
with $G$ and $\lambda_\ast$ being given by
\begin{eqnarray}
    \label{eqn:equili_sin}
    \sin\frac{\lambda_\ast}{2} &=& \sin\frac{\psi}{2} \sin\frac{\tilde{\lambda}}{2} =
    2\sqrt{RT} \sin\frac{\psi}{2} \sin \frac{\lambda}{2}\\
    \label{eqn:equili_g}
    G &=& \frac{1}{2\pi^2}\ln t_f\xi.
\end{eqnarray}

The logarithmic terms in (\ref{eqn:fes_noise}) are connected with
the Fermi Edge Singularity found in metals.\cite{Mahan67,ND69} It is
interesting to note that the form (\ref{eqn:equili_sin}) includes
the well-known result for the Anderson orthogonality catastrophe
problem in a single lead \cite{Anderson67} as a special case. The
quantity of interest is the overlap $\langle0|0'\rangle$ between
$|0\rangle$, the ground state of an unperturbed metal and
$|0'\rangle$, the wavefunction of the same metal at a time $t_f$
after switching on a (core hole) potential. In the simplest case,
the potential can be well described by a single energy-independent
phase shift which is exactly equivalent to the phase $\psi$, with
$e^{i\psi/2} = B + iA$ associated with the modulated barrier we have
been considering. The overlap can be written as
$\langle0|0'\rangle=\langle0|e^{i\psi(t)}|0\rangle=\prod_j\cos\frac{\alpha_j}{2}$,
where the $\frac{\alpha_j}{2}$ are the eigenvalues $e^{\pm\alpha_j}$
of $he^{i\psi/2}he^{-i\psi/2}$. This is equal to
$\sqrt{\chi_{2}(\lambda=\pi,T=\frac{1}{2})}$. With
$\lambda=\pi,T=\frac{1}{2}$, $\lambda_*= \psi$ and we recover
Anderson's result \cite{Anderson67}:
$\langle0|e^{i\psi(t)}|0\rangle=(\xi t_f)^{-\psi^2/(4\pi^2)}$.

\section{General Case}\label{sec:general}

In the  general case, which is equivalent to having both a bias
voltage between the leads and a time-dependent profile for the
barrier, we are not aware of the existence of a simple relation
mapping the problem onto an equivalent bias voltage problem.
However, (\ref{eqn:ivanov}) allows for  the calculation of the FCS
as long as the spectrum of the density matrix of outgoing states is
available.\cite{Abanov08} We can diagonalize $n^{out}$ in
(\ref{eqn:nout}) numerically to find the eigenvalues $n_j$ and use
(\ref{eqn:ivanov}) to compute the FCS.

Numerically, it is convenient to work in the energy domain with
a discrete energy spectrum. (An alternative is to compute
the dynamics of the system  directly. For a finite system the spectrum is then
automatically truncated and
  the determinant is properly regularized\cite{Schonhammer07,Schonhammer09}.)
We introduce a periodic boundary condition
in time, with period $t_p$, discretizing the energy spectrum of the
scattering matrix with an energy separation $\omega_0=2\pi/t_p$. By
choosing $t_p\ll t_f$ we can compute the counting statistics with
large number of cycles, giving the characteristic function as
$\chi(\lambda)\approx[\chi_0(\lambda)]^{t_f/t_p}$, where
$\chi_0(\lambda)$ is the FCS for single period.
We can also set $t_p\gg t_f$ and study the behavior of a device when
operated in one-shot mode.

Fourier transformed into the energy domain, the individual matrix
elements of $A(t)$ and $B(t)$ are
$\mathbf{X}_{mn}=\frac{1}{t_p}\int_0^{t_p}e^{-i(m-n)\omega_0t}X(t)dt$
where $X$ stands for $A$ or $B$. The neighboring rows of $X_{mn}$
have the same elements though they are shifted from each other by
one column. If $X(t)$ is sufficiently smooth, by which we mean its
Fourier transform decays faster than $\omega_0^\nu$, with $\nu<-1$.
We can cut off the Fourier series and limit the approximated
summation within $|m-n|<M$ with $M$ chosen large enough to achieve the desired
accuracy. The truncated matrices $\mathbf{A}$ and $\mathbf{B}$ are
blockwise tridiagonal with each block size $M\times M$. After some
manipulation, the diagonalization of the infinite dimensional matrix
$n^{out}$ is approximately equivalent to the diagonalization of a
$2M$ dimensional matrix in energy space. Details of this procedure
are summarized in App. \ref{sec:app0}.

In this section, we look at cases where the direct diagonalization
of the matrix $n^{out}$ allows the study of the FCS of a tunnel
barrier with time-dependent scattering amplitudes with a bias
voltage applied between the leads. We concentrate, in particular, on
the example of a barrier used as a low noise source and study how
the noise and degree of entanglement are affected by deviations from
the optimal pulses which have been
proposed.\cite{Sherkunov09,Keeling08} Mostly, we will study barrier
and voltage profiles which are either combinations (or close to
combinations) of the quantized Lorentzian pulses. These can be
applied as voltage pulses to a lead with (\ref{eqn:lorentzian})
describing the corresponding Faraday flux $e^{i\psi(t)}$, or with
(\ref{eqn:lorentzian}) describing the barrier profile $e^{i\phi(t)}
= B(t)+iA(t)$.

\subsection{Quantized Pulses}\label{ssec:op}

To operate a tunneling barrier as a single electron source at low
temperatures requires a voltage pulse which excites a single
electron excitation in one lead. This can be achieved by creating a
MES with a single Lorentzian pulse applied between the leads. At low
temperatures, the noise produced by such a device comes from two
sources: shot and equilibrium noise. \cite{BlaB2000} The shot noise
for the simplest case of a barrier with constant transmission
probability, $T$, is proportional to $T(1-T)$ for a single MES
pulse. This suggests that one should aim to open the barrier fully
($T=1$) to increase the chance of single electron emission. However,
if we open the barrier in an arbitrary way, the equilibrium noise
becomes important. A solution \cite{Sherkunov09} is to
combine the creation of the MES, choosing a bias voltage pulse
giving Faraday flux
\begin{equation}
e^{i\psi(t)}= \frac{t-t_0-i\tau_0}{t-t_0+i\tau_0} \label{eqn:mes0}
\end{equation}
in the incoming states, with a carefully-chosen opening profile for
the barrier which minimizes the total noise.

\begin{figure}[t]
    \begin{center}
        \includegraphics[width=0.3\textwidth]{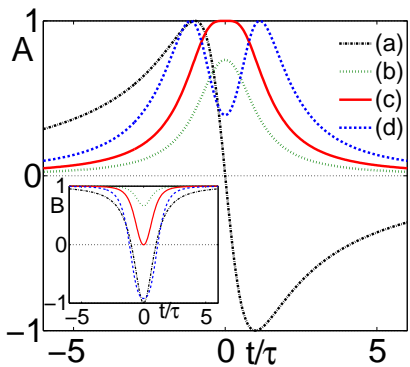}
    \end{center}
    \caption{\label{fig:mep}
    (Color online) Transmission coefficients $A(t)$ for some opening profiles for a barrier.
    The centers of the pulses are at $t=0$, with time scaled in units of
        the pulse width $\tau$.
    (a) Single Lorentzian pulse (see \ref{eqn:entangler}) for optimal entangler.
    (b) Two Lorentzian pulses, with opposite polarities (see \ref{eqn:mep_source})
    with $\tau_1=\tau_2=\tau$ and pulse separation $t_1/\tau\approx0.6$.
    (c) Same as in (b) with separation $t_1/\tau\approx0.83$ (optimal).
    (d) Same as in (b) with $t_1/\tau\approx1.2$.
    The inset shows the corresponding reflection amplitude $B$ as a function of $t/\tau$.
    Only in cases (b) and (c) do both $A$ and $B$ remain positive during the pulse lifetime.}
\end{figure}

Since the profile generating the maximal entanglement
(\ref{eqn:entangler}), as illustrated in Fig. \ref{fig:mep}(a),
fully opens the barrier twice without any equilibrium noise
contribution depending logarithmically on $t_f$, a first guess is
that it might also work as a possible profile for the opening of the
barrier when used as a single electron source. However, the barrier
opening scheme based on (\ref{eqn:entangler}) is not ideal for a
single electron emission. Though the logarithmic term in the
equilibrium noise is absent, the profile does generate background
noise with (see \ref{eqn:cf_barrier}) $\langle\langle
Q^2\rangle\rangle_0=\frac{\partial^2\log\chi}{\partial(i\lambda)^2}
=\sum_j^N\frac{1}{2}\sin^2\frac{\alpha_j}{2}\leq\frac{N}{2}$. For
the profile (\ref{eqn:entangler}) and $N=1$ and $\alpha=\pi$ we find
$\langle\langle Q^2\rangle\rangle=\frac{1}{2}$ which is actually the
maximum for a single pulse.

We consider instead a pair of such pulses:
\begin{equation}
    e^{i\phi(t)}=\frac{t-t_1/2-i\tau_1}{t-t_1/2+i\tau_1}\frac{t+t_1/2+i\tau_2}{t+t_1/2-i\tau_2},
    \label{eqn:mep_source}
\end{equation}
where the separation of the two pulses (with widths $\tau_1$ and
$\tau_2$ respectively) is $t_1$. A little algebra shows that $A(t)$
and $B(t)$ do not change sign only if $\tau_1=\tau_2$ and
$\tau_1/t_1 \ge\frac{1}{2}+\frac{1}{\sqrt{2}}$. Corresponding
typical profiles are shown in Fig. \ref{fig:mep}(b-c). The ratio
$\tau_1/t_1=\frac{1}{2}+\frac{1}{\sqrt{2}}$ gives a transparency of
the barrier which has a single maximum with $A=1$ at $t=0$, as shown
in Fig. \ref{fig:mep}(c). We have found empirically that this
separation gives the best combination (low total noise and highest
probability for the transfer of one electron). Exciting a single
excitation in the incoming states of one lead and coordinating the
timing of this excitation with the opening of the barrier, allows the
single particle excitation in the incoming states to impinge upon the
barrier when it is fully open. An advantage of the profile
(\ref{eqn:mep_source}) is that, owing to cancelation between the two
components (at $t=-t_1/2$ and $t=t_1/2$) at long times, the
transmission amplitude $A(t)$ approaches zero faster than for the
profile (\ref{eqn:entangler}). In addition, the noise generated by
this opening profile $\langle\langle
Q^2\rangle\rangle_0\approx0.23$, which is less than half that
generated by (\ref{eqn:entangler}). We expect that the profile
(\ref{eqn:mep_source}), with an optimally chosen ratio for
$t_1/\tau_1$, should be a good candidate for designing an on-demand
single electron source at ultralow temperatures.

\begin{figure}[t]
    \begin{center}
        \includegraphics[width=0.45\textwidth]{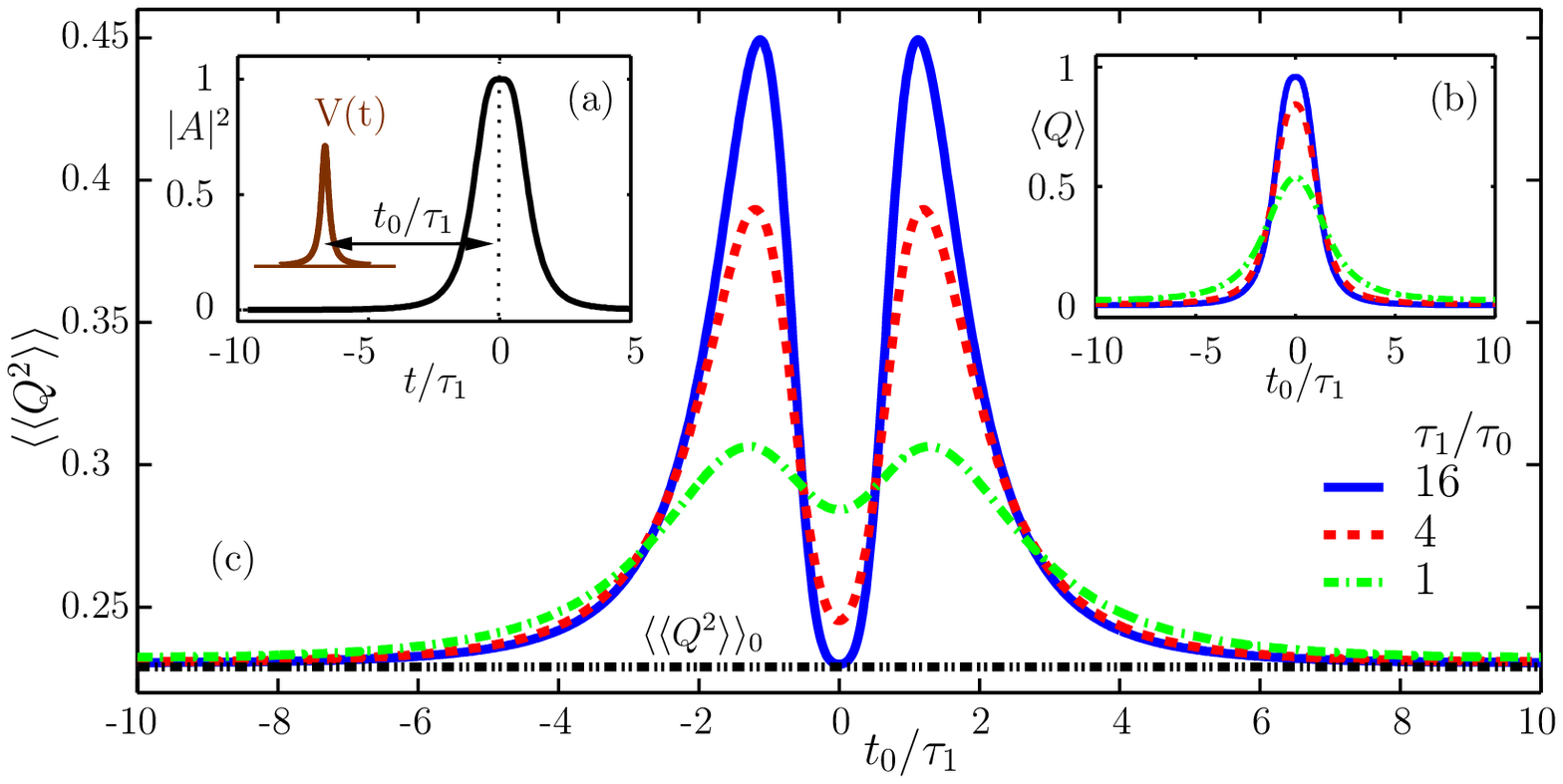}\\
        \includegraphics[width=0.45\textwidth]{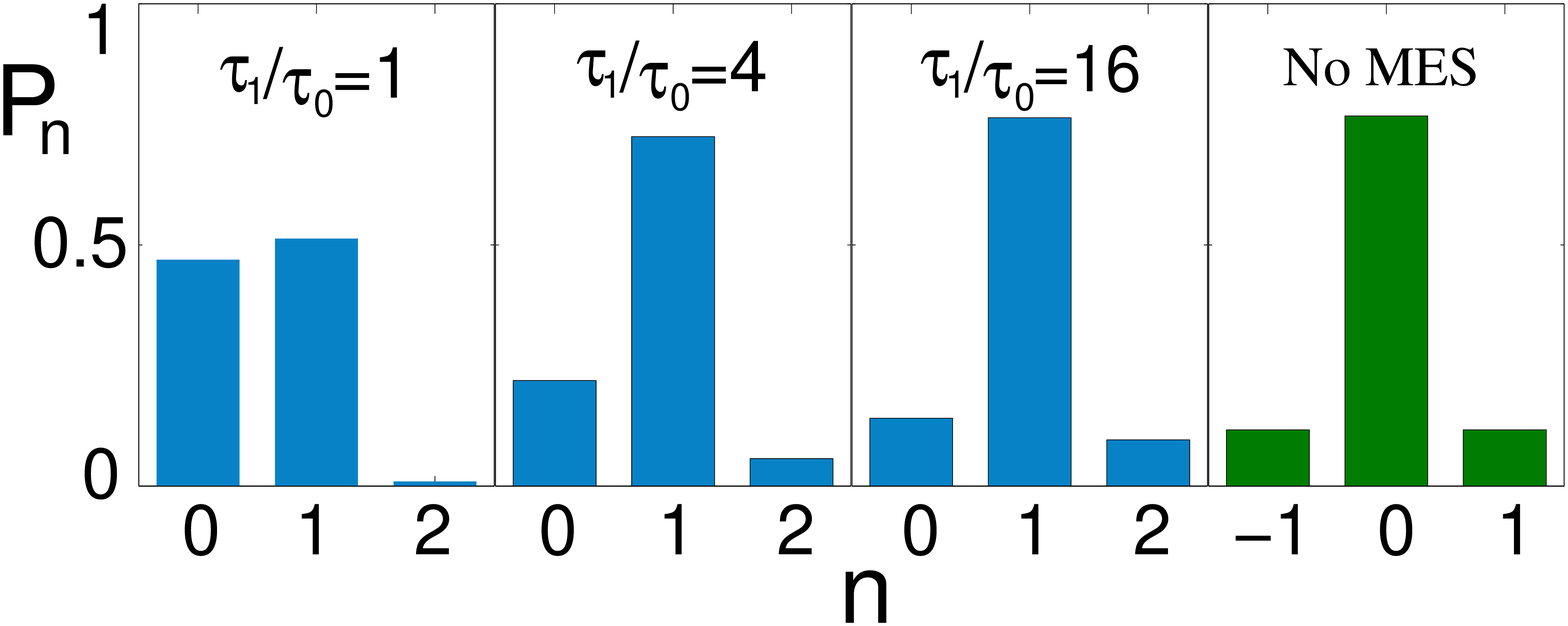}
    \end{center}
    \caption{\label{fig:source}
    (Color online) Upper panel: Single unidirectional event in a quantum contact
    with tunable transparency (numerical calculations).
    (a) Transparency of the barrier as a function of time given by (\ref{eqn:mep_source}).
    We also show an MES applied to a lead (\ref{eqn:mes0}).
    (b) Average charge transfer $\langle Q\rangle$ between the two
    leads as a function of $t_0$.
    The maximum of $\langle Q\rangle$ is around $0.96$ for the narrowest MES considered $\tau_1/\tau_0=16$.
    (c) Noise in a biased contact with transparency
    (\ref{eqn:mep_source})
    as a function of $t_0/\tau_1$. The minimal value of the noise
    corresponds to $t_0=0$ and, for the narrowest MES pulse ($\tau_1/\tau_0=16$),
    it almost drops to the value due to barrier modulation:
    $\langle \langle Q^2 \rangle \rangle_{_0}$ (dotted line).
    The two maxima in the noise occur when $t_0/\tau_1 \approx 1$
    and the transparency is almost $1/2$.
    Lower panel: Probability distribution $P_n$ for $\tau_1/\tau_0=1$,$4$ and $16$ at
    $t_0=0$, as well as the probability distribution generated by profile (\ref{eqn:mep_source})
    only without the MES applied on the lead.
    We see that in the limit $\tau_1/\tau_0\rightarrow\infty$, $P_n$ approaches $P^0_{n\pm1}$.
    (See text)}
\end{figure}

For $\tau_1/t_1=\frac{1}{2}+\frac{1}{\sqrt{2}}$ and with a single
pulse $e^{i\psi}=\frac{t-t_0-i\tau_0}{t-t_0+i\tau_0}$ applied to the
lead, we have diagonalized the density matrix $n_{out}$ numerically.
Fig. \ref{fig:source}(a) shows results for the noise $\langle
\langle Q^2\rangle \rangle$ in the system as a function of the
separation $t_0$ between the center of the MES pulse and the maximum
of barrier transparency (for which $A=1$ at $t=0$). The maximum
values of the noise occurs when $t_0 \approx \tau_1$ and the
transparency coefficient is almost $1/2$. This is the regime where
the barrier is acting as a $50\%$ beam-splitter. The minimum of
$\langle \langle Q^2\rangle \rangle$ corresponds to $t_0=0$, when
$|A|^2=1$. At $t_0=0$ the transferred charge as well as the quality
factor $\langle Q\rangle/\langle\langle Q^2\rangle\rangle$ attains
its maximum. Fig. \ref{fig:source}(b) shows the average transferred
charge $\langle Q\rangle$ as a function of $t_0$. If the pulse
applied to the lead is narrow compared to the opening time of the
barrier ($\tau_0 \ll \tau_1$), the minimum value of the noise is
essentially set by the noise associated with the opening of the
barrier, namely $\langle \langle Q^2 \rangle \rangle_{_0}$. The transferred
charge approaches $1$ when $\tau_0/\tau_1\rightarrow0$ and the
probability distribution for transferred charge, $P_n$, approaches
$P^0_{n\pm1}$ (the sign depends on the polarity of the voltage
pulse applied to the incoming states in one of the leads) is shown in
the lower panel in Fig \ref{fig:source}. Here $P^0_{n\pm 1}$ is the
probability distribution for charge transfer associated with the
opening of the barrier without a bias pulse applied between the leads.

\begin{figure}[!ht]
    \begin{center}
        \includegraphics[width=0.3\textwidth]{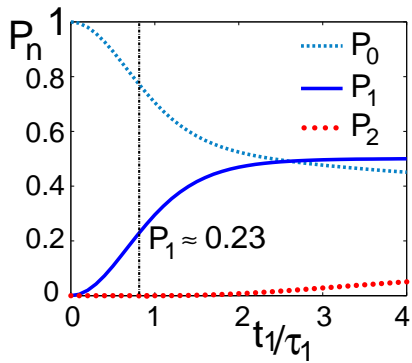}
    \end{center}
    \caption{\label{fig:ent_src}
    (color online) Evolution of the probability for electron-hole pair emission as a function
of pulse separation $t_1/\tau_1$ for a barrier profile
(\ref{eqn:mep_source}) with $\tau_2=\tau_1$. The dashed line gives
the probability that nothing happens ($P_0$). The solid line is the
probability of generating a single entangled electron-hole pair
($P_1$). The dotted line show the probability for two particle-hole
pair emission ($P_2$). The vertical cut is the position where the
single electron source profile (\ref{eqn:mep_source}) operates.}
\end{figure}

Fig \ref{fig:source} shows that, when $\tau_1/\tau_0=1$, the emission of
2 electrons is strongly suppressed. If it was important to have a source
which only emitted single electrons and there was a method for discarding
`non-events' in which no electron was emitted, this protocol would work
well. The
suppression of double electron emission can be understood as follows.
The width of pulse determines the energy profile of the
excited particles or holes. When $\tau_0\sim\tau_1$, the particle
excitation induced by the changing barrier profile has the same energy
profile as that of the incoming excitation induced by the bias voltage
applied between the leads. As the two pulses are coincident in time
the Pauli exclusion
principle leads to destructive interference between the two excitations
in one lead and either only one particle or no particle will be
transmitted through the QPC as a result.

The profile (\ref{eqn:mep_source}) does not involve a change of sign
of either $A(t)$ or $B(t)$ and, consequently, should be easier to
implement experimentally than the profile (\ref{eqn:entangler}). We
therefore consider how well the profile (\ref{eqn:mep_source}) might
work as a (non-optimal) electronic entangler. Fig. \ref{fig:ent_src}
shows the probability of electron-hole pair production $P_n$ versus
pulse separation $t_1/\tau_1$. The vertical line corresponds to
$t_1/\tau_1=1/(\frac{1}{2}+\frac{1}{\sqrt{2}})\approx0.83$, beyond
which a sign change in $A(t)$ is necessary and the experimental
implementation is expected to be more involved. We see that the
emission of two electron-hole pairs is very unlikely. The only two
significant outcomes are: no excitation, created with probability
$P_0$, and the creation of a single (entangled) electron hole pair
which is created with probability $P_1$. The corresponding
entanglement entropy at this point is $S=P_1S_{Bell}\approx0.23$,
{\it ie\/} just under half the theoretical maximum.

\subsection{Non-Quantized Pulses}\label{ssec:nonop}
So far, we have discussed (combinations of) quantized Lorentzian
pulses applied to a lead (see \ref{eqn:lorentzian}) and/or barrier
opening profiles (see \ref{eqn:mep_source}) that excite a
bounded number of electron-hole pairs without the accompanying
equilibrium noise which grows logarithmically with $t_0$.
Here, we focus on the increased
noise resulting from deviations from the ideal quantized
pulses.

We consider the case of a modulated barrier, with the
following non-quantized opening profile:
\begin{equation}
    B(t) +iA(t) = e^{i\phi(t)}=\left(\frac{t-i\tau_1}{t+i\tau_1}\right)^{\gamma_1}
    \left(\frac{t-t_0-i\tau_2}{t-t_0+i\tau_2}\right)^{\gamma_2}.
    \label{eqn:nonop1}
\end{equation}
where $\gamma_{1,2} \in \mathbb{R}$. We look at the simplest case
$\gamma_2=1-\gamma_1$ and $\gamma_2=-\gamma_1$ and choose $\tau_1 =
\tau_2=\tau$. In both cases $t_0$ plays the role of the measurement
time $t_f$, provided $t_0\gg\tau$. The profile (\ref{eqn:nonop1})
with $t_0\gg \tau$ describes a barrier with a transmission
amplitude, $A$, which changes from $0$ to  $\sin 2 \pi \gamma$
within a period of $\tau$ around $t=0$ and closes at $t=t_0$. There
will then be a contribution to the noise which increases
logarithmically with $t_0$ (this is just the equilibrium noise
contribution). The mapping between the case of a barrier profile and
a voltage bias across a junction with constant transparency means
that this profile also models non-quantized voltage pulses applied
between the leads. This mapping includes a doubling of the total
phase change $2\phi = \psi$ (see Sec. \ref{ssec:barrier}), which
means that (\ref{eqn:nonop1}) describes two voltage pulses with
Faraday flux $4\pi \gamma_{1,2}$. If the $\gamma_i$ are not both
integer or half-integer, there is a net phase shift of
$2\pi\gamma_1$ and $2\pi(\gamma_1 + \gamma_2)$ between the rotated
and the unperturbed states with consequent FES effects.

The FCS for the profile (\ref{eqn:entangler}),
which gives the optimum level of entanglement,
are equivalent to those for \emph{two} quantized Lorentzian pulses, when the problem is
mapped to the case of a junction with a bias, because of the associated
doubling of the phase $2\phi = \psi$,
This suggests that the pulse (\ref{eqn:entangler}) applied to the barrier
may be a special case of two separate pulses, each of which corresponds
to a single quantized pulse in the case of a bias between the leads.
In particular, a pulse
\begin{equation}
    e^{i\phi(t)} =
    \left(\frac{t-i\tau_1}{t+i\tau_1}\right)^{\frac{1}{2}}
    \left(\frac{t-t_0-i\tau_2}{t-t_0+i\tau_2}\right)^{\frac{1}{2}},
    \label{eqn:mep_root}
\end{equation}
consists of two pulses centered at $0$ and $t_0$ with
widths $\tau_1$ and $\tau_2$. The absence
of a logarithmic contribution to the noise in this case should be
expected, because the barrier is closed at all times except for
a period $\tau$ around $0$ and $t_0$.
The effect of pulses (\ref{eqn:mep_root}) on the states of the
system can be found by working in the basis in which the scattering
matrix in (\ref{eqn:smat}) is diagonal. With $A$ and $B$ both real
($\psi=0$ as there is no applied bias), the scattering matrix is
diagonal for all times in the basis $c_{1,2}=(c_L\pm ic_R)/\sqrt{2}$
and its eigenvalues as a function of time are given by the values of
$e^{\pm i \phi(t)}$ in (\ref{eqn:mep_root}). The two components of
the pulse centered on $t=0$ and $t=t_0$ induce phase shifts of
$\pm \pi$ in each channel and are individually as far as possible from
quantized pulses. However their effect on the FCS of the charge
transfer between the left and right leads depends only on the
difference in phase shift between the two channels and each of the two
components therefore contribute to the FCS as quantized pulses.

\begin{figure}[t]
    \begin{center}
\includegraphics[height=0.22\textheight]{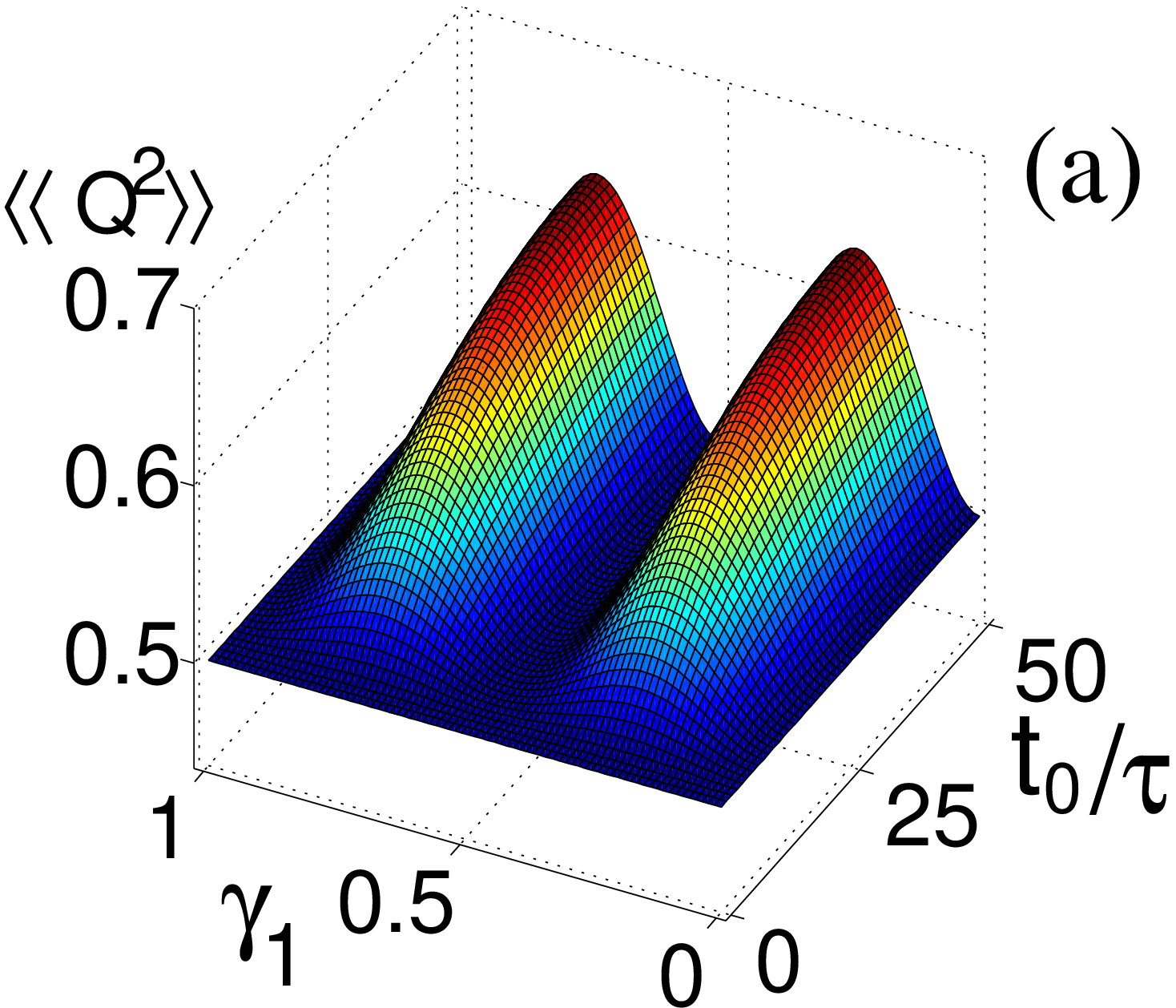}\\
\includegraphics[height=0.22\textheight]{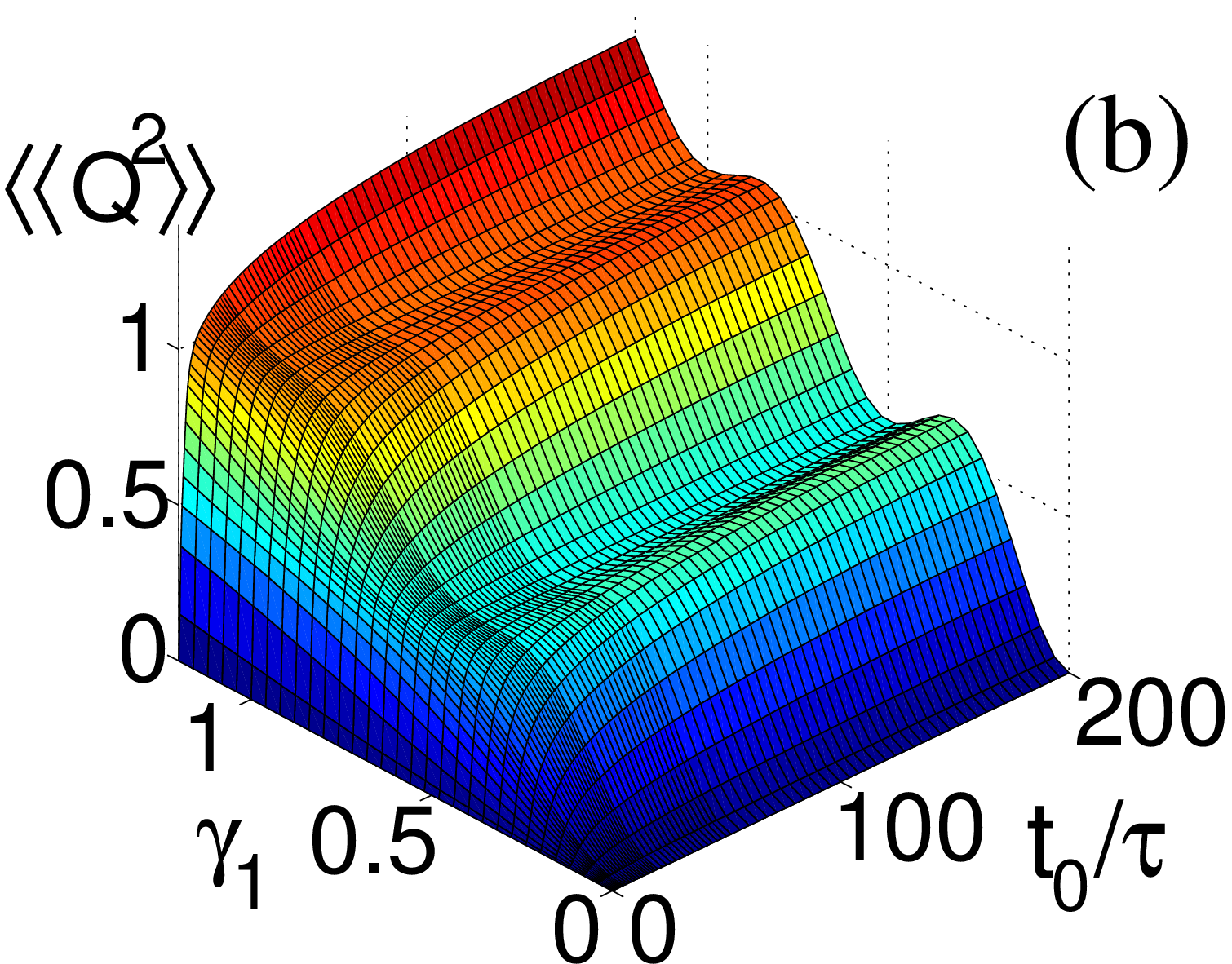}
    \end{center}
    \caption{\label{fig:nonop1}
    (Color online) The noise produced in the charge transfer across an unbiased barrier when its profile is varied
    according to (\ref{eqn:nonop1}). Results are shown as a function of the separation, $t_0/\tau$, between the
    two non-quantized pulses at $t=0$ and $t=t_0$, and of
    the exponent $\gamma_1$ with (a) $\gamma_2=1-\gamma_1$ and
    (b) $\gamma_2=-\gamma_1$. In case (a) the values $\gamma_1=0.5$ and $\gamma_1=1$
    correspond to quantized pulses (when mapped to the problem of a biased lead, there is a doubling
    of the total phase change). The noise for all other values of $\gamma_1$ increases logarithmically
    with $t_0$. This is the equilibrium noise contribution. In case (b) the noise $\gamma_1=0.5$ and
    $\gamma_1=1$ saturates when the pulse separation $t_0$ is much larger than the
    pulse widths $\tau$. For all other $\gamma_1$ there is again the equilibrium noise contribution which
    grows logarithmically with measurement time $t_0$.
}
\end{figure}

The noise generated by the opening profile (\ref{eqn:nonop1}) with
$\tau_1=\tau_2 = \tau$ is shown in Fig. \ref{fig:nonop1}(a) for the
case $\gamma_2=1-\gamma_1$ and in Fig.  \ref{fig:nonop1}(b) for the
case $\gamma_2=-\gamma_1$. In both these cases the penalty for
missing quantization of the pulses is small. The increase in noise
at fixed $t_0$, as $\gamma_1$ deviates from integer or half-integer
values, is slow. (For example, with
$\gamma_1=1/2$ and $t_0/\tau=50$, a $10$\% deviation in $\gamma_1$
introduces additional noise of only of $2$\% of the quantized value.)
This suggests that any
reasonable experimental implementation of such pulses should allow
the exploitation we have described (as entanglers or as electron
sources). Case (a) corresponds to two Lorentzian (non-quantized)
pulses with the same polarity separated by a time $t_0$. There are
two peaks, the height of which grow logarithmically with $t_0$, with
a flat valley in between. When $\gamma_1=1/2$ the profile is that of
(\ref{eqn:mep_root}), and we find, as expected, results equivalent
to the profile (\ref{eqn:entangler}), {\it ie} a total noise equal
to 1/2 and independent of $t_0$. For other values of $\gamma_1$ the
noise grows with a logarithmic dependence on $t_0$. When $\tau<t_0$,
the problem is that of two equivalent FES problems: The effect of
the first pulse is to give rise to scattering phase shifts of $2\pi
\gamma_1$ and $-2\pi\gamma_1$ in the two independent channels in
which the scattering matrix $S(t)$ is diagonal. In case (b) there
are two oppositely polarized (non-quantized) Lorentzian pulses. At
small pulse separation, $t_0$, the two pulses partially cancel and the
noise is low. In this limit the barrier transparency, $T(t)$ remains
close to zero and vanishes when $t_0=0$. Saturation of the noise at
some finite value  for large $t_0/\tau$ occurs only when
$\gamma_1=1/2$ and $1$, when the two components of the pulse
contribute to the FCS and each corresponds to quantized MES in the
equivalent lead problem. For all other values of $\gamma_1$, we
again find noise which grows with the logarithmic dependence with
$t_0$ expected for a FES.

Another issue with the quantized Lorentzian pulses, used either as
voltage pulses or to open the barrier, is that they are defined over
an infinite time interval. In practice, the long tail behavior will
be restricted to some closed interval $[-t_0, t_0]$ and this cut-off
is likely to introduce additional noise. We have found, however,
that amending the profile at points $\pm
t_0$  and appending an exponential tail
$A_e(t)=A(t_0)e^{-|t-t_0|/\epsilon}$ to model this effect
that there is virtually no difference between this trimmed profile
and the ideal profile. This suggests that such
deviations from ideal pulses are unlikely to affect the operation of
devices in this regime.

Finally, we return to the problem introduced in Sec.
\ref{ssec:fes}, where two delta pulses (see \ref{eqn:delta_functions}) with
opposite polarities were applied to one incoming channel.
We argued that
the FCS could be written as the product of two factors (\ref{eqn:chi1chi2}).
One factor, $\chi_1$,
is associated
with the sudden jump in scattering phase shift and the other, $\chi_2$,
with the FES problem. To
compute $\chi_1(\lambda)$, we need to specify the exact shape of the Dirac
delta function. Here we use a Lorentzian in
the limit of vanishing width $\tau\rightarrow0$:
\cite{Reason02}
\begin{equation}
    V(t) = \frac{\psi}{2\pi}\left(\frac{\tau}{t^2+\tau^2}-\frac{\tau}{(t-t_0)^2+\tau^2}\right).
    \label{eqn:nonop_mes}
\end{equation}
The FCS are fully determined by the spectrum of $n^{out}$, or
equivalently, the eigenvalues $e^{\pm i\alpha_j}$ of $h\tilde{h}$ in
the form $\sin^2\frac{\alpha_j}{2}$. We note that this problem is
equivalent to a limiting case of the non-quantized barrier profile
(\ref{eqn:nonop1}) shown in Fig. \ref{fig:nonop1}(b): After mapping,
the barrier profile in (\ref{eqn:nonop1}) has $\gamma_1
=-\gamma_2=\frac{\psi}{4\pi}$ and $\tau \rightarrow 0$.

In the following, we show that the factorization of the counting
statistics which we propose (\ref{eqn:chi1chi2}) is apparent in the
separation of the values of $\sin^2\frac{\alpha_j}{2}$. We rewrite
$\psi=2\pi k+\psi_0$, where $k\in \mathbb{Z}$ and
$\psi_0\in[0,2\pi)$. When $\psi_0=0$, from (\ref{eqn:vnb}) the
counting statistics are described by $k$ bi-directional events with
eigenvalues such that $\sin^2\frac{\alpha_j}{2}=1$ with $j=1,\ldots,
k$. Hence $\chi_1(\lambda)=[1-4RT\sin^2 \frac{\lambda}{2}]^k$. On
the other hand, for states which contribute to (\ref{eqn:chi2_fes}),
they are rotated with eigenvalues $e^{\pm i \alpha_j}$
such that $\sin^2\frac{\alpha_j}{2}$ is small (but total number of
such eigenvalues is large). For the case when $\psi_0\ne0$, we have
verified numerically that  the  $k$ pairs of eigenvalues of
$h\tilde{h}$ with $e^{\pm i\alpha_j}$ remain. In addition there is
one pair of eigenvalues associated with $\psi_0$ which gives $\sin^2
\frac{\alpha_0}{2}$ in the range $[0,1)$. Here $\alpha_0/2$ is the
rotation angle associated with the phase $\psi_0$. As a result, the
total FCS (\ref{eqn:app_sin}) should be well approximated by the
form:
\begin{eqnarray}
\nonumber
\chi(\lambda)&=&\prod_{j=0}^\infty\left(1-4RT\sin^2\frac{\alpha_j}{2}\sin^2\frac{\lambda}{2}\right)\\
\nonumber
&\approx&(1-4RT\sin^2\frac{\alpha_0}{2}\sin^2\frac{\lambda}{2})
\\ &\cdot&\left(1-4RT\sin^2\frac{\lambda}{2}\right)^k\cdot
\chi_2(\lambda). \label{eqn:chi_fac}
\end{eqnarray}
This value of $\alpha_0$ can be computed
numerically once for a simple case like $k=0$, and tabulated over
the range $\psi_0\in[0,2\pi)$.

\begin{figure}[t]
    \begin{center}
        \includegraphics[width=0.43\textwidth]{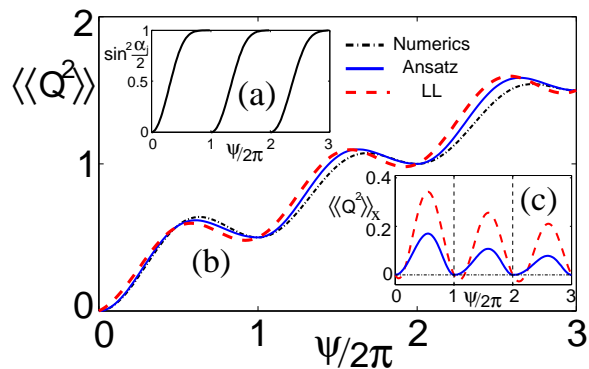}
    \end{center}
    \caption{\label{fig:nonop2} (Color online)
    (a) Evolution of the extra eigenvalue $\sin^2\frac{\alpha_0}{2}$ with $\psi$
     in (\ref{eqn:delta_functions}) .
    (b) Noise computed from three approaches: Numerics (dash dotted),
    ansatz (\ref{eqn:ans_noise}) (solid) and LL's result (\ref{eqn:fes_noise}) (dashed).
    (c) $\langle\langle Q^2\rangle\rangle_x$ (see text)
    computed for LL's result and ansatz (\ref{eqn:ans_noise}).}
\end{figure}

To test above idea, as well as to obtain the unknown eigenvalue
$\sin^2\frac{\alpha_0}{2}$, we compute the FCS and the values
$\sin^2\frac{\alpha_j}{2}$ for (\ref{eqn:nonop_mes}) numerically for
different total phase shift $\psi$ in (\ref{eqn:delta_functions}).
The pulse separation is fixed at $t_0/\tau=500$ and the contact
transparency is $T=1/2$. With cut-off energy $\xi$ as the only
fitting parameter, the ansatz (\ref{eqn:chi_fac}) agrees with the
numerical result for $\chi(\lambda)$ in the parameter space
$\{\psi,\lambda\}$ everywhere with a maximum difference of 3\%. This
agreement justifies our decomposition of $\chi(\lambda)$
based on the separation of eigenvalues of $n^{out}$. The evolution
of $\sin^2\frac{\alpha_0}{2}$ with $\psi$ is drawn in Fig.
\ref{fig:nonop2}(a) and echoes that reported for a tunnel junction
driven by a sinusoidal voltage as a function of the voltage
amplitude.\cite{Vanevic2008} As $\psi$ increases, the value
$\sin^2\frac{\alpha_0}{2}$ increases gradually and saturates at $1$.

The noise can be computed using (\ref{eqn:chi_fac}) and is
\begin{equation}
\langle\langle Q^2\rangle\rangle =
    2RT[\frac{2}{\pi^2}\sin^2\frac{\psi_0}{2}\ln t_0\xi+k+\sin^2\frac{\alpha_0}{2}].
\label{eqn:ans_noise}
\end{equation}
In Fig. \ref{fig:nonop2}(b) we show the noise computed from three
approaches: i) exactly computed numerically from (\ref{eqn:ivanov});
ii) using out ansatz (\ref{eqn:ans_noise}) and iii) from the
expansion used by LL (\ref{eqn:fes_noise}). The cut-off energy $\xi$
is used as fitting parameter to give the best agreement with the
exact result. The similarity between the noise produced here and the
noise generation in Fig. \ref{fig:nonop1}(b) for fixed $t_f$ can be
understood using the mapping between the problem we are considering
here of an applied bias voltage and that of a modulate barrier
profile discussed in Sec. \ref{ssec:barrier}. In both cases the
contribution proportional to the $\log t_0\xi$ is suppressed at
quantized phase shift $\psi$.

The differences between the two approximate treatments and the
exact results are associated with the ``large'' eigenvalues. We
define the quantity $\langle\langle Q^2\rangle\rangle_x
=\langle\langle Q^2\rangle\rangle_{nu}-2RTx(\psi)$, with
$x(\psi)=\frac{\psi}{2\pi}$ for the expansion (\ref{eqn:fes_noise})
and $x(\psi)=k+\sin^2\frac{\alpha_0}{2}$ for our ansatz
(\ref{eqn:chi_fac}).  $\langle\langle Q^2\rangle\rangle_{nu}$ is the
exact value of the noise computed numerically. According to
(\ref{eqn:fes_noise}) as well as (\ref{eqn:ans_noise}),
$\langle\langle Q^2\rangle\rangle_x$ is proportional to
$\sin^2\frac{\psi}{2}$, and, if the cut-off energy $\xi$ is
independent of $\psi$, is periodic. It should also vanish at
$\psi=0,2\pi,4\pi\ldots$. We show $\langle\langle
Q^2\rangle\rangle_x$ for the two cases respectively in Fig.
\ref{fig:nonop2}(c). We see that the ansatz (\ref{eqn:ans_noise})
works qualitatively correctly while the expansion
(\ref{eqn:fes_noise}) gives the wrong positions for the minima. However, we find
that
$\langle\langle Q^2\rangle\rangle_{nu}$ is not strictly periodic (the
amplitude of the oscillation decreases). We attribute this to  the fact that
the separation into the two factors $\chi_1$ (corresponding to $k$ large values
of $\sin^2\alpha_j$) and $\chi_2$ (only small values) is not complete
and that the value $\sin^2\alpha_{k+1}/2$ can be significant particularly for
small $k$.

\section{Conclusion}\label{sec:conclusion}
We have discussed the Full Counting Statistics (FCS) of the charge
transferred across a quantum point contact. We have illustrated the
power of a mapping between the case of a biased barrier with
constant transmission and reflection amplitudes and the case of a
barrier with time-dependent profile but no bias. With this mapping
we have showed that known results for the two cases, which had been
previously obtained using different, and generally involved
calculations, can be understood using the basis of quantized
Lorentzian pulses\cite{IvaLL97} or Minimal Excitation
States\cite{KeelingKlichLev06} (MES). Examples include the optimal
protocol for electron entanglement, \cite{Sherkunov09} the FCS of a
sinusoidally driven barrier, \cite{AndreevKamenev2000} and the Fermi
Edge Singularity. \cite{ND69}

For the purposes both of conceptual understanding and computation,
we have argued that the problem is simplest when approached through
the eigenvalues of the density matrix of the outgoing states in one
of the leads. For the general case, which corresponds to applying
both a bias and varying the barrier profile with time, we have
developed a numerical scheme for computing exactly the FCS for a
device and used this to compute the FCS for a tunnel barrier
operated as an electron source. We have also studied how the
deviation from an ideal pulse affect the quality of operation of a
device with low noise or an entangler. We showed that the noise
levels were remarkably insensitive to deviations from quantized
Lorentzian pulses associated with the long-time behavior. For
deviations away from the quantization of the pulses in the case of a
modulation of barrier profile, we found that, as expected, this led
to the reappearance of the equilibrium noise contribution, which
increases logarithmically with barrier opening time (see Fig.
\ref{fig:nonop1}).

This work is supported by EPSRC-GB (Contract No. EP/D065135/1).

\appendix
\section{Numerical Diagonalization of (\ref{eqn:nout})}\label{sec:app0}
We describe the numerical procedure for diagonalizing $n^{out}$ in
(\ref{eqn:nout}). We need a discretized finite dimensional
expression for $\mathbf{XnX}^\dagger$, where $\mathbf{X}$ stands for
the transmission and reflection matrices $\mathbf{A}$ and
$\mathbf{B}$ in frequency space. $\mathbf{n}$ is an infinite
dimensional diagonal matrix with elements
$n_{mn}=\delta_{mn}\theta(-m+\epsilon)$, where $\epsilon<\omega_0$,
to avoid $m=0$ in the $\theta$-function. The matrix $\mathbf{X}$ is
blockwise tridiagonal:
\begin{equation*}
    \mathbf{X} = \left(\begin{array}{ccccc}
        \ddots  & \ddots       & \ddots       & \mbox{}      & \mbox{} \\
        \mbox{} & \mathbf{X}_2 & \mathbf{X}_0 & \mathbf{X}_1 & \mbox{} \\
        \mbox{} & \mbox{}      & \ddots       & \ddots       & \ddots
    \end{array}\right),
\end{equation*}
where $\mathbf{X}_{0,1,2}$ is square matrix with truncated dimension
$M$ (see main text). $\mathbf{X}_1$ ($\mathbf{X}_2$) is lower
(upper) triangular matrix with null diagonal elements.

Exploiting the property $B(t)B^\ast(t)+A(t)A^\ast(t)=1$, which in
frequency space states $\mathbf{B}\mathbf{B}^\dagger
+\mathbf{A}\mathbf{A}^\dagger=\mathbf{I}$ ($\mathbf{I}$ is the
infinite dimensional identity matrix), we find the following
properties for the finite block sub-matrices with dimension $M$:
$\sum_{i=0}^2(\mathbf{B}_i\mathbf{B}_i^\dagger+\mathbf{A}_i\mathbf{A}_i^\dagger)=\mathbf{I}$,
$\mathbf{B}_2\mathbf{B}_0^\dagger+\mathbf{B}_0\mathbf{B}_1^\dagger
+\mathbf{A}_2\mathbf{A}_0^\dagger+\mathbf{A}_0\mathbf{A}_1^\dagger=\mathbf{0}$,
and
$\mathbf{A}_1\mathbf{A}_2^\dagger=\mathbf{B}_1\mathbf{B}_2^\dagger
=\mathbf{0}$.

Compute $n^{out}$ in (\ref{eqn:nout}) in frequency space and remove
the part equivalent to $n^{in}$ at low and high energies, we arrive
at a much simplified form of $n^{out}$ suitable for numerical
diagonalization
\begin{equation}
    \mathbf{n}^{out} = \left(\begin{array}{cc}
        \mathbf{B}_1\mathbf{B}_1^\dagger+\mathbf{A}_1\mathbf{A}_1^\dagger &
        \mathbf{B}_1\mathbf{B}_0^\dagger+\mathbf{A}_1\mathbf{A}_0^\dagger \\
        \mathbf{B}_0\mathbf{B}_1^\dagger+\mathbf{A}_0\mathbf{A}_1^\dagger &
        \mathbf{I}-\mathbf{B}_2\mathbf{B}_2^\dagger-\mathbf{A}_2\mathbf{A}_2^\dagger
    \end{array}\right).
    \label{eqn:app_nout}
\end{equation}
The sought-after spectrum, $n_j$, is obtained by diagonalizing
(\ref{eqn:app_nout}) directly.

\section{Riemann-Hilbert solution to the FCS for barrier
opening at non-zero temperature} \label{sec:app1}

Here we give a brief derivation of the FCS for the pure opening
problem (region 2, Sec. \ref{ssec:fes}) within the Riemann-Hilbert
approach at finite temperature. The scattering matrix $S(t)$ in
(\ref{eqn:chi2_fes}) takes the form
\begin{equation*}
    S(t)=\left(\begin{array}{cc}
        \cos\frac{\phi}{2} & \sin\frac{\phi}{2}\\
        -\sin\frac{\phi}{2} & \cos\frac{\phi}{2}
    \end{array}\right),~\mbox{for t}\in[0,t_f]
\end{equation*}
and equals the identity otherwise. Following the notation in Ref.
\onlinecite{MA03,dAMuz05}, we introduce the matrix $R(\lambda)=
Se^{i\lambda L}S^\dagger e^{-i\lambda L}$, with
$L=\begin{pmatrix}1&0\\0&0\end{pmatrix}$.  The characteristic
function $\ln\chi$ reads
\begin{equation}
    \ln\chi(\lambda)=\mbox{Tr}[n\ln R] + \mbox{Tr}[\ln(1-n+nR)-n\ln R]
    \label{eqn:app_lnx}
\end{equation}
where $\mbox{Tr}$ operates on both time and channel space. $n$ is
the fermionic operator at finite temperature $1/\beta$:
\begin{equation*}
n(t,t')=\frac{i}{2\pi}\frac{\pi/\beta}{\sinh(\pi(t-t')/\beta+i0)}.
\end{equation*}

Since there is no average charge transfer, the first term of
(\ref{eqn:app_lnx}), $\mbox{Tr}[n\ln R]=i\lambda \langle Q\rangle$,
is equal to zero. The FCS is given by the second term,
\begin{equation*}
\ln\chi(\lambda)=\frac{i}{2\pi}\int_0^\lambda d\lambda\int
dt~\mbox{tr}
    \left[\frac{d\ln Y_+}{dt}\frac{d\ln R}{d\lambda}\right],
\end{equation*}
where $\mbox{tr}$ is a trace over channel space only. The
Riemann-Hilbert solution $Y(z)$ is a bounded matrix-valued analytic
function in the strip $-\beta/2<\mbox{Im}z<\beta/2$ except along
the cut $z\in[0,t_f]$, where $Y_-(t)Y_+^{-1}(t)=R(\lambda)$
$Y_+(t)\equiv Y(t+i0)$ and $Y_-(t)\equiv Y(t-i0)$ (see Ref.
\onlinecite{MA03,dAMuz05,Braunecker06}). Making the substitution
$\sin\frac{\lambda_\ast}{2}=\sin\frac{\phi}{2}\sin\frac{\lambda}{2}$,
$R(\lambda)$ can be diagonalized in a time-independent basis with
eigenvalues $e^{\pm i\lambda_\ast}$. Using explicitly the finite
temperature solution of the RH problem,\cite{Braunecker06}
\begin{equation*}
Y_+(t)=\exp\left(\frac{1}{2\pi i}\int_0^{t_f}dt' \frac{\cosh (\pi
t/\beta)}{\cosh (\pi t'/\beta)}\frac{\pi\ln R/\beta}{\sinh
(\pi(t-t')/\beta)}\right)
\end{equation*}
we obtain
\begin{equation*}
\ln\chi(\lambda)=-\frac{\lambda_\ast^2}{2\pi^2}\ln\left(\frac{\sinh(\pi
t_f/\beta)}{\sinh(\pi\xi^{-1}/\beta)}\right),
\end{equation*}
with $\xi$ as a cut-off. Taking the zero temperature limit
$\beta\rightarrow \infty$, we arrive at the characteristic function
(\ref{eqn:chi2_fes}) along with definitions (\ref{eqn:equili_sin})
and (\ref{eqn:equili_g}).

\bibliographystyle{apsrev}
\bibliography{fcs}

\begin{thebibliography}{42}
\bibitem[{Rea()}]{Reason02}
\bibinfo{note}{We have also looked at the case where a Gaussian pulse
  $V(t)=\frac{\phi}{4\pi\sqrt{\pi\tau}}(e^{-t^2/4\tau}-e^{-(t-t_0)^2/4\tau})$
  is used to approximate the delta function as $t_0/\tau\rightarrow\infty$. The
  result doesn't show any significant difference.}
\end{thebibliography}

\end{document}